\documentclass[aps,showpacs, prb,twocolumn,superscriptaddress]{revtex4}%
\usepackage{epsfig,dsfont,amssymb,amsmath,amsthm,amsfonts,amsbsy,mathrsfs}
\usepackage{graphicx}
\usepackage{amsmath}
\usepackage{amssymb}
\begin{document}
\title{Structures, stability, mechanical and electronic properties of $\alpha$-boron and its twined brother $\alpha$*-boron}
\author{Chaoyu He}
\affiliation{Hunan Key Laboratory for Micro-Nano Energy Materials
and Devices, Xiangtan University, Hunan 411105, P. R. China.}
\affiliation{Laboratory for Quantum Engineering and Micro-Nano
Energy Technology, Faculty of Materials and Optoelectronic Physics,
Xiangtan University, Hunan 411105, P. R. China}
\author{Jianxin Zhong}
\email{jxzhong@xtu.edu.cn}\affiliation{Hunan Key Laboratory for
Micro-Nano Energy Materials and Devices, Xiangtan University, Hunan
411105, P. R. China.} \affiliation{Laboratory for Quantum
Engineering and Micro-Nano Energy Technology, Faculty of Materials
and Optoelectronic Physics, Xiangtan University, Hunan 411105, P. R.
China}
\date{\today}
\begin{abstract}
The structures, stability, mechanical and electronic properties of
$\alpha$-boron and its twined brother $\alpha$*-boron have been
studied by first-principles calculations. Both $\alpha$-boron and
$\alpha$*-boron consist of equivalent icosahedra B$_{12}$ clusters
in different connecting configurations of "3S-6D-3S" and "2S-6D-4S",
respectively. The total energy calculations show that
$\alpha$*-boron is less stable than $\alpha$-boron but more
favorable than $\beta$-boron and $\gamma$-boron at zero pressure.
Both $\alpha$-boron and $\alpha$*-boron are confirmed dynamically
and mechanically stable. The mechanical and electronic properties of
$\alpha$-boron and $\alpha$*-boron indicate that they are potential
superhard semiconducting phases of element boron. \\
\end{abstract}
\maketitle \indent Boron is considered as the most mystic element
due to its fascinating chemical complexity that leads to vast
variety of polymorphs \cite{B1, B2, B3, B4, B5, B6, B7}. Among these
polymorphs, rhombohedral $\alpha$-boron and $\beta$-boron (with 12
and 106 atoms in unit cells, respectively) are well-known and almost
degenerate in energy at ambient condition. $\alpha$-boron is more
favorable than $\beta$-boron at high pressure and $\beta$-boron is
more stable in high temperature condition. In 2008, Oganov et
al.\cite{B8, B9, B10} discovered an ionic phase of elemental boron
(orthorhombic $\gamma$-boron with 28 atoms in its unit cell), which
consists of icosahedral B$_{12}$ clusters and interstitial B$_2$
pairs acting as anions and cations, respectively, in a NaCl-type
arrangement. This new phase of boron is stable between 19 and 89 GPa
but quenchable at ambient conditions, which confirms the previous
discovery of a new crystal of boron \cite{B7} and thus provides us
the missing piece of a puzzle of the phase diagram of boron \cite{B9}.\\
\indent Born atoms in its element crystals prefer forming
icosahedral B$_{12}$ cluster, where each boron atom connects five
neighbors with intra-B$_{12}$ B-B bonds. In $\alpha$-boron,
equivalent icosahedral B$_{12}$ clusters connect to each other
directly with inter-B$_{12}$ B-B bonds and each of them has 12
equivalent neighbouring B$_{12}$ clusters. In $\gamma$-boron, each
icosahedral B$_{12}$ connects to ten equivalent neighbouring
B$_{12}$ clusters directly with inter-B$_{12}$ B-B bonds and two
secondly adjacent B$_{12}$ clusters indirectly by interstitial B$_2$
pairs. The connectivity between icosahedral
B$_{12}$ clusters and interstitial atoms in $\beta$-boron becomes more complex. \\
\indent $\alpha$-boron is not the only phase of boron consisting of
pure icosahedral B$_{12}$ clusters (without any interstitial atoms
like those in $\beta$-boron and $\gamma$-boron). Very recently,
Pickard et al. obtained \cite{B11} a new meta-stable phase of boron
by an ab initio random structure searching method. This new phase
belongs to Cmcm space group and can be considered as a twined
polymorph of $\alpha$-boron with new connectivity (different from
that in $\alpha$-boron boron) between icosahedral B$_{12}$ clusters
which is 0.01 eV/atom less stable than $\alpha$-boron and more
favorable than $\beta$-boron. With remarkable energetic stability,
such a phase of boron is a potential superhard material. But there
is no further discussion about the structure, dynamical stability,
mechanical and electronic properties of this new phase. In this
paper, based on the density functional theory (DFT) method, we
systematically study the structure, dynamical and mechanical
stability, mechanical and electronic properties of this new boron
phase (we call it $\alpha$*-boron in the following) and compare them with those of $\alpha$-boron. \\
\section{Computational Details}
\indent All calculations are carried out using the density
functional theory (DFT) with general gradient approximation (GGA)
\cite{gga} as implemented in Vienna ab initio simulation package
(VASP) \cite{22, 23}. The interactions between nucleus and the
valence electrons are described by the projector augmented wave
(PAW) method \cite{24, 25}. A plane-wave basis with a cutoff energy
of 500 eV is used to expand the wave functions of all the systems
considered in present work. The Brillouin Zone (BZ) sample meshes
for all systems are set to be denser enough (less that 0.2 1/{\AA})
in our calculations. Crystal lattices and atoms positions for every
systems are fully optimized (under different external pressure) up
to the residual force on each atom less than 0.005 eV/{\AA} through
the conjugate-gradient algorithm. The vibrational properties of
$\alpha$-boron and $\alpha$*-boron are studied by using the phonon
package \cite{26} with the forces calculated from VASP. For the
mechanical properties calculations, different numbers of deformed
$\alpha$-boron, $\alpha$*-boron and $\gamma$-boron are
considered according to their crystal symmetries.\\
\begin{figure}
\includegraphics[width=3.5in]{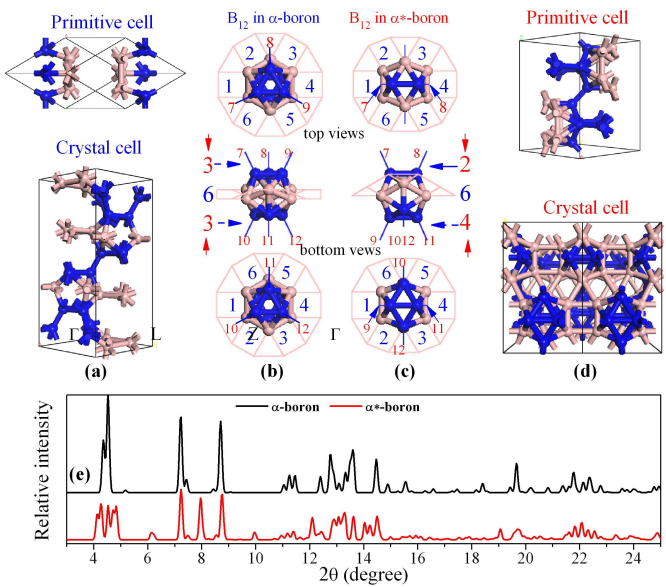}\\
\caption{Primitive view and crystal view of $\alpha$-boron (a);
Sketch maps showing the structural characters of $\alpha$-boron (b)
and $\alpha$*-boron (c); Primitive view and crystal view of
$\alpha$*-boron (d); Simulated X-ray ($\lambda$= 0.31851 {\AA})
diffraction patterns of $\alpha$-boron and $\alpha$*-boron (e). In
figures the blue and brown balls indicate the S-type and D-type
boron atoms, respectively.}\label{fig1}
\end{figure}
\section{Results and Discussions}
\subsection{Structures}
\indent The optimized structures of $\alpha$-boron and
$\alpha$*-boron in both primitive cell (top) and crystal cell
(bottom) are shown in Fig.~\ref{fig1} (a) and (d), respectively. The
hexagonal crystal cell of $\alpha$-boron belongs to R-3m (166) space
group. Under zero pressure, its lattice constants are a=b=4.9 {\AA}
and c=12.55 {\AA}. Two inequivalent boron atoms B1 and B2 in
$\alpha$-boron locate at positions of (0.237, 0.119, 0.108) and
(0.197, 0.803, 0.024), respectively. In $\alpha$-boron, each B1 atom
possesses five intra-B$_{12}$ B-B bonds and single (S-)
inter-B$_{12}$ B-B bond (S-type, colored in blue). Each B2 atom
possesses five intra-B$_{12}$ B-B bonds and double (D-)
inter-B$_{12}$ B-B bonds (D-type, colored in brown). The
orthorhombic $\alpha$*-boron belongs to Cmcm (63) space group. Its
equilibrium lattice parameters are a=4.88 {\AA}, b=8.85 {\AA} and c=
8.06 {\AA}, respectively. There are five inequivalent boron atoms
B1, B2, B3, B4 and B5 in $\alpha$*-boron. They locate at positions
of (0.667, 0.334, 0.75), (0.818, 0.507, 0.75), (1.000, 0.563,
0.924), (0.797, 0.169, 0.639) and (0.500, 0.265, 0.568),
respectively. Similar to those in $\alpha$-boron, the five
inequivalent boron atoms in $\alpha$*-boron can be divided into two
types: S-type where the three boron atoms (B1, B2 and B3) colored in
blue have five intra-B$_{12}$ B-B bonds and single (S-)
inter-B$_{12}$ B-B bonds and D-type where the two brown boron atoms
(B4 and B5) possess five intra-B$_{12}$ B-B bonds and double (D-)
inter-B$_{12}$ B-B bonds. Although $\alpha$-boron and
$\alpha$*-boron contain different numbers of inequivalent boron
atoms, both structures contain only one inequivalent icosahedral
B$_{12}$ cluster. That is the reason we refer them as twined
structures of element boron.\\
\indent In both $\alpha$-boron and $\alpha$*-boron, each icosahedral
B$_{12}$ cluster possesses twelve neighboring B$_{12}$ clusters
distributing in different stacking manners. In $\alpha$-boron,
twelve neighboring B$_{12}$ clusters distribute around the center
B$_{12}$ cluster with a "3S-6D-3S" connecting configuration. As
indicated in Fig~\ref{fig1} (b), three top-position B$_{12}$
clusters (indicated as 7, 8 and 9) connect to the center B$_{12}$
cluster with single inter-B$_{12}$ B-B bonds formed by S-type boron
atoms, six middle-position B$_{12}$ clusters (indicated as 1, 2, 3,
4, 5 and 6) connect to the center B$_{12}$ clusters with double
inter-B$_{12}$ B-B bonds formed by D-type boron atoms and three
bottom-position B$_{12}$ clusters (indicated as 10 ,11 and 12)
connect to the center B$_{12}$ cluster with single inter-B$_{12}$
B-B bonds formed by S-type boron atoms, forming a "3S-6D-3S"
configuration. In $\alpha$*-boron, the twelve neighboring B$_{12}$
clusters distribute around the center B$_{12}$ cluster with a
"2S-6D-4S" connecting configuration as indicated in Fig~\ref{fig1}
(c). Two top-position B$_{12}$ clusters (indicated as 7 and 8)
connect to the center B$_{12}$ cluster with single inter-B$_{12}$
B-B bonds formed by S-type boron atoms, six middle-position B$_{12}$
clusters (indicated as 1, 2, 3, 4, 5 and 6) connect to the center
B$_{12}$ cluster with double inter-B$_{12}$ B-B bonds formed by
D-type boron atoms and four bottom-position B$_{12}$ clusters (9,
10, 11 and 12) connect to the center B$_{12}$ cluster with
inter-B$_{12}$ B-B bonds formed by S-type boron atoms. In both
$\alpha$-boron and $\alpha$*-boron, S-type atoms connect to D-type
atoms only with intra-B$_{12}$ B-B bonds and never meet each other
in different B$_{12}$ clusters.\\
\indent Both $\alpha$-boron and $\alpha$*-boron contain equivalent
icosahedral B$_{12}$ clusters in their crystal structure. The
different connecting configurations of "3S-6D-3S" and "2S-6D-4S" for
$\alpha$-boron and $\alpha$*-boron result in different space group
of R-3m (166) and Cmcm (63), respectively. The different structures
of $\alpha$-boron and $\alpha$*-boron also result in different X-ray
diffraction patterns. Fig.~\ref{fig1} (e) shows the X-ray
diffraction patterns of $\alpha$-boron and $\alpha$*-boron. We can
see that almost all the diffraction peaks of $\alpha$-boron can be
found in the XRD of $\alpha$*-boron with reduced intensities, such
as the peaks at 4.5$^\circ$, 7.2$^\circ$ and 8.6$^\circ$. In
addition, $\alpha$*-boron possesses characteristic diffraction peaks
forbidden in $\alpha$-boron, which locate at 4.7$^\circ$,
6.2$^\circ$, 8.0$^\circ$, 9.9$^\circ$, 12.09$^\circ$, 19.05$^\circ$
and 23.33$^\circ$. These characteristic peaks are helpful for experimental
identification of  $\alpha$*-boron.\\
\begin{table*}
  \centering
  \caption{The calculated values of elastic constants C$_{ij}$ (GPa), bulk modulus B (GPa), shear modulus G (GPa),
  Vicker's hardness H${_v}$ (GPa) of the $\alpha$-boron, $\alpha$*-boron and $\gamma$-boron.}\label{tabI}
\begin{tabular}{c c c c c c c c c c c c c c}
\hline \hline
Systems         &C$_{11}$   &C$_{12}$   &C$_{13}$   &C$_{14}$   &C$_{22}$   &C$_{23}$    &C$_{33}$    &C$_{44}$   &C$_{55}$   &C$_{66}$   &B        &G       &H${_v}$\\
\hline
$\alpha$-boron    &454.64    &109.86     &45.46      &20.60      &   -       &  -        &605.21     &208.03     & -          &  -        &212.38   &201.63  &38.96\\
$\alpha$*-boron   &466.22    &53.99      &77.49      &  -        &569.69     &56.67      &444.13     &211.73     &122.05      &217.76     &205.58   &190.91  &36.60\\
$\gamma$-boron    &604.58    &80.26      &37.65      &  -        &534.42     &80.70      &448.36     &243.20     &223.34      &253.00     &218.79   &235.26  &50.12\\
\hline \hline
\end{tabular}
\end{table*}
\begin{figure}
\includegraphics[width=3.5in]{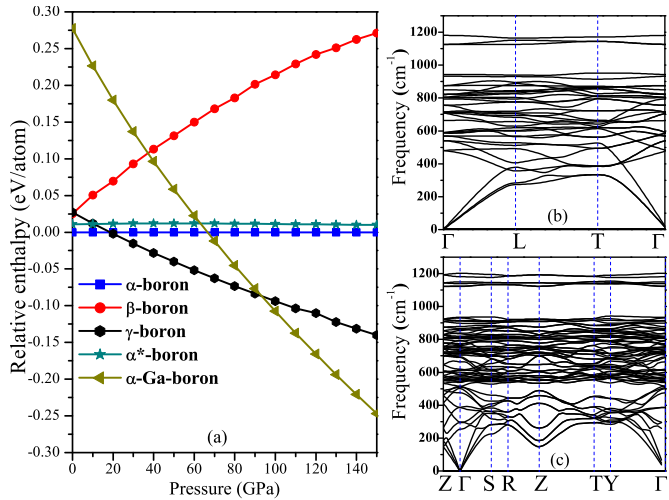}\\
\caption{The enthalpy curves for $\alpha$*-boron, $\gamma$-boron,
$\beta$-boron and $\alpha$-Ga-boron as functions of pressure
relative to that of $\alpha$-boron (a); The simulated phonon band
structures of $\alpha$-boron (b) and $\alpha$*-boron
(c).}\label{fig2}
\end{figure}
\subsection{stability}
\indent In this part, we discuss the energetic and dynamical
stability of $\alpha$-boron and $\alpha$*-boron. In Fig.~\ref{fig2}
(a) we show the enthalpy curves of $\alpha$*-boron, $\gamma$-boron,
$\beta$-boron and $\alpha$-Ga-boron as functions of pressures
relative to that of $\alpha$-boron. We can see that $\alpha$*-boron
possesses remarkable energetic stability comparable to that of its
twined brother $\alpha$-boron at any pressure range. Its enthalpy is
only about 0.01 eV/atom above that of $\alpha$-boron. At zero
pressure, both $\alpha$-boron and $\alpha$*-boron are more favorable
than $\gamma$-boron, $\beta$-boron and $\alpha$-Ga-boron. When the
external pressure is increased, they transform to $\gamma$-boron at
20 GPa and then to $\alpha$-Ga-boron at 94 GPa. Our results are in
good agreement with previous first-principles calculations \cite{B9}.\\
\indent The remarkable stability indicates that $\alpha$*-boron has
high probability to coexist with $\alpha$-boron as the ground state
of element boron in the pressure range from 0-21 GPa. But how about
the possibility? Is it a dynamically stable phase? We need to know
the vibrational properties of such a new crystalline structure of
boron to answer the question. We investigate the phonon properties
of $\alpha$-boron and $\alpha$*-boron by phonon package \cite{26}
with interatomic forces calculated from vasp \cite{22, 23}. The
calculated phonon band structures of $\alpha$-boron and
$\alpha$*-boron are shown in Fig.~\ref{fig2} (b) and (c),
respectively. We can see that there are no imaginary frequencies in
both $\alpha$-boron and $\alpha$*-boron, indicating that they are
dynamically stable. Above discussions about the energetic and
dynamical stability of $\alpha$*-boron indicate that it is a
promising meta-stable phase of element boron.\\
\subsection{Mechanical properties}
\indent The elastic constants of $\alpha$-boron and $\alpha$*-boron
are calculated as the second-order coefficient in the polynomial
function of distortion parameter $\delta$ used to fit their total
energies according to the Hooke's law. In view of their differences
in crystal symmetries, six and nine groups of deformations are
applied on $\alpha$-boron and $\alpha$*-boron ($\gamma$-boron) in
our calculations, respectively. The elastic constants of
$\alpha$-boron and $\alpha$*-boron are summarized in Table I with
those of $\gamma$-boron used for comparison. Our calculated elastic
constants for $\alpha$-boron and $\gamma$-boron are in good
agreement with those in previous first-principle calculations
\cite{com1, com2}. Based on these results, we can see that the six
independent elastic constants of $\alpha$-boron satisfy the
corresponding mechanical stability criteria\cite{mec1, mec2, mec3}
for a hexagonal crystal with C$_{66}$=(C$_{11}$-C$_{12}$)/2$>$0,
C$_{11}$+C$_{12}$-2(C$_{13}$)$^2$/C$_{33}$$>$0, and C$_{44}$$>$0.
The orthorhombic $\alpha$*-boron possesses nine independent elastic
constants. As listed in Table I, these elastic constants also
satisfy the corresponding mechanical stability criteria \cite{mec1,
mec2, mec3} for a orthorhombic crystal with
C$_{11}$+C$_{12}$-C$_{21}$$>$0, C$_{11}$+C$_{33}$-2C$_{13}$$>$0,
C$_{22}$+C$_{33}$-2C$_{23}$$>$0, C$_{11}$$>$0, C$_{22}$$>$0,
C$_{33}$$>$0, C$_{44}$$>$0, C$_{55}$$>$0, C$_{66}$$>$0, and
C$_{11}$+C$_{22}$+C$_{33}$+2C$_{12}$+2C$_{13}$+2C$_{23}$$>$0.\\
\indent The bulk modulus (\textbf{B}) and shear modulus (\textbf{G})
are evaluated according to Hill's formula \cite{Hill} based on the
calculated elastic constants. The calculated \textbf{B} and
\textbf{G} for $\alpha$*-boron are 205.58 GPa and 190.91 GPa,
respectively, which are very close to those of $\alpha$-boron. To
further analyze the hardness of $\alpha$-boron and $\alpha$*-boron,
we adopt the recently introduced empirical scheme \cite{hard} to
evaluate their Vicker's hardness (H${_v}$) determined their
\textbf{B} and \textbf{G} as
\textbf{H${_v}$=2(G${^3}$/B${^2}$)${^{0.585}}$-3}. The calculated
values of Vicker's hardness for $\alpha$-boron and $\alpha$*-boron
are 38.96 GPa and 36.60 GPa, respectively. These values indicate
that both $\alpha$-boron and $\alpha$*-boron are potential superhard phases of element boron.\\
\begin{figure}
\includegraphics[width=3.50in]{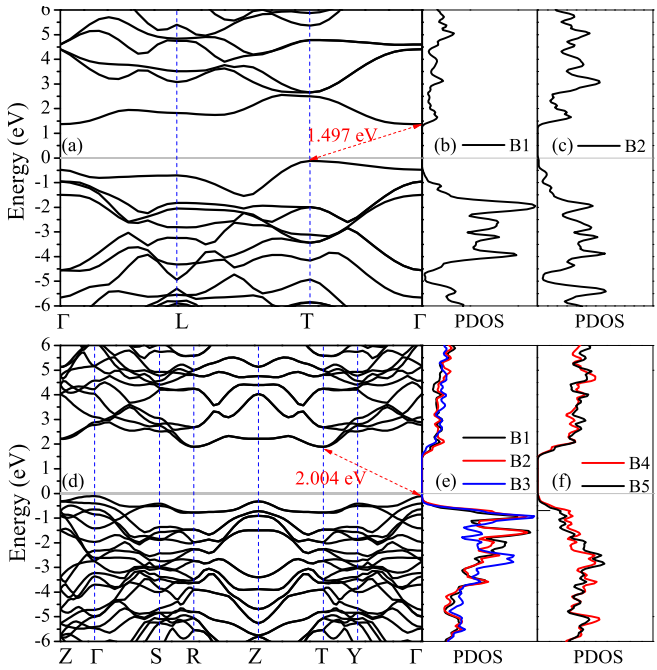}
\caption{The band structures of $\alpha$-boron (a) and corresponding
local density of state projected on the S-type (b) and the D-type
(c) boron atoms. The band structures of $\alpha$*-boron (d) and
corresponding local density of state projected on the S-type (e) and
the D-type (f) boron atoms.}\label{fig3}
\end{figure}
\begin{figure}
\includegraphics[width=3.50in]{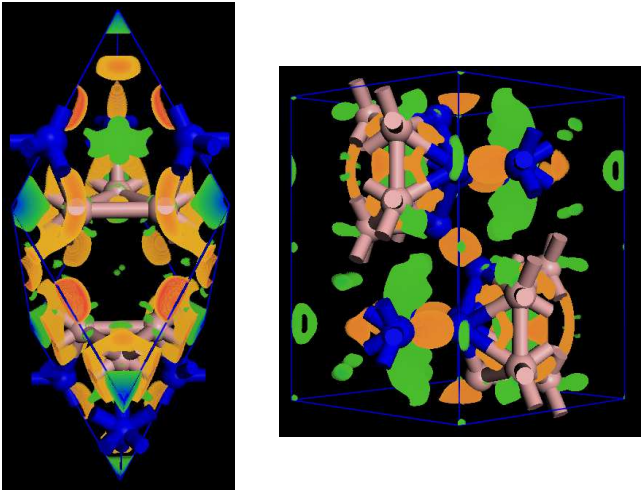}
\caption{The charge difference density for $\alpha$-boron (left) and
$\alpha$*-boron (right). Electrons transfer from the green-blue area
to the red-yellow area.}\label{fig4}
\end{figure}
\subsection{Electronic properties}
\indent Both $\alpha$-boron and $\alpha$*-boron contain pure
icosahedral B$_{12}$ clusters with different connecting
configurations. Such a structural difference may result in different
electronic properties in $\alpha$-boron and $\alpha$*-boron. In
Fig.~\ref{fig3} (a) and (d), we show the band structures of
$\alpha$-boron and $\alpha$*-boron, respectively. From the band
structures, we can see that both of them are indirect-band-gap
semiconductors. Differently, the valence band maximum (VBM) and
conduction band minimum (CBM) in $\alpha$-boron locate at T-point
(0.5, 0.5, 0.5) and $\Gamma$-point (0.0, 0.0, 0.0), respectively,
separated by a forbidden band with a gap of 1.497 eV. But the VBM
and CBM in $\alpha$*-boron locate at $\Gamma$-point (0.0, 0.0, 0.0)
and T-point (0.5, 0.5, 0.5), respectively, forming a indirect-band
gap of 2.004 eV. Based on the local density of states (LDOS)
projected on S-type and D-type boron atoms as shown in
Fig.~\ref{fig3}, we can see that the VBM states in both
$\alpha$-boron and $\alpha$*-boron are contributed from the D-type
boron atoms and their CBM states are derived from S-type boron
atoms. In both $\alpha$-boron and $\alpha$*-boron the whole valence
band is mainly contributed by the S-type boron atoms and the whole
conduction band is mainly contributed by the D-type boron atoms.
These results indicate that the valence electrons density mainly
comes from the S-type boron atoms and the hole density mainly comes from the D-type atoms.\\
\indent The corresponding charge difference density (CDD) of
$\alpha$-boron and $\alpha$*-boron is shown in Fig.~\ref{fig4} to
confirm and understand the electronic properties discussed above.
From the CDD of $\alpha$-boron shown in the left in Fig.~\ref{fig4},
we can see that the D-type boron atoms (brown balls) loss electrons
and the S-type boron atoms (blue balls) gain electrons. The CDD of
$\alpha$*-boron is shown in the right panel in Fig.~\ref{fig4}.
Similar to that of $\alpha$-boron, the D-type boron atoms (brown
balls) in $\alpha$*-boron loss electrons and the S-type boron atoms
(blue balls) gain electrons. These CDD results reveal that the
bonding electrons in both $\alpha$-boron and $\alpha$*-boron prefer
to distribute on the inter-B$_{12}$ B-B bonds formed by the S-type
boron atoms than those formed by the D-type boron atoms and the
intra-B$_{12}$ B-B bonds.
\section{Conclusion}
\indent  We have studied the structure, stability, mechanical and
electronic properties of a new orthorhombic boron phase
$\alpha$*-boron and compared them with those of its twined brother
$\alpha$-boron. Both $\alpha$-boron and $\alpha$*-boron consist of
equivalent icosahedra B$_{12}$ clusters with different connecting
configurations of "3S-6D-3S" and "2S-6D-4S", respectively. Our
results show that $\alpha$*-boron is a dynamically and mechanically
viable phase of element boron. The remarkable energetic stability
and considerable hardness of 36.8 GPa of $\alpha$*-boron indicate
that it is a promising superhard material with potential applications in industry.\\
\section{Acknowledgements}
This work is supported by the National Natural Science Foundation of
China (Grant Nos. 51172191 and 11074211), the National Basic
Research Program of China (2012CB921303), the Program for New
Century Excellent Talents in University (Grant No.NCET-10-0169), and
the Scientific Research Fund of Hunan Provincial Education
Department (Grant Nos. 09K033, 10K065, 10A118).

\end{document}